\documentclass[aps,prl,twocolumn,showpacs]{revtex4}

\usepackage{dcolumn}
\usepackage{amssymb}
\usepackage{graphicx}

\begin{document}

\newcommand \be {\begin{equation}}
\newcommand \ee {\end{equation}}
\newcommand \bea {\begin{eqnarray}}
\newcommand \eea {\end{eqnarray}}
\newcommand \nn {\nonumber}
\newcommand \ve {\varepsilon}

\title{Definition and relevance of nonequilibrium intensive thermodynamic
parameters}
\author{Eric Bertin,$^{1}$ Olivier Dauchot,$^2$ and Michel Droz$^1$}
\affiliation{$^1$Department of Theoretical Physics, University of Geneva, CH-1211 Geneva 4, Switzerland\\
$^2$SPEC, CEA Saclay, F-91191 Gif-sur-Yvette Cedex, France}
\date{\today}

\begin{abstract}
We show that intensive thermodynamic parameters associated
to additive conserved quantities can be naturally defined from a statistical
approach in far-from-equilibrium steady-state systems,
under few assumptions, and without any detailed balance requirement.
It may apply, e.g., to dissipative systems like granular gases where volume
or mass is still conserved, or to systems with periodic boundary
conditions where fluxes of conserved quantities are present.
We emphasize the usefulness of this concept to characterize the coexistence
of different nonequilibrium phases, and discuss the influence of the contact
between two different systems, in relation with measurement issues.
\end{abstract}


\maketitle

One of the essential features of intensive thermodynamic
parameters (ITPs), like temperature, pressure or chemical potential,
is that these parameters take equal values in two
different systems that can exchange the conjugated extensive quantity
(energy, volume, or particles) once equilibrium is reached.
This is a strong property, since it holds even if the systems put
into contact have very different microscopic dynamics.
This equilibration between different systems
makes ITPs a powerful concept to describe the influence of the environment
or phase coexistence, and is essential to allow, say, the temperature of
a system to be measured with a thermometer.

Several approaches have been developed in
nonequilibrium statistical physics to generalize the notion of ITP,
usually focusing on temperature.
For aging systems, an effective temperature may be defined from the long-time
slope of the fluctuation-dissipation relation (FDR)
\cite{CuKuPe,Kurchan,Ritort}.
For systems weakly driven into a stationary nonequilibrium
state, ITPs are defined locally as in equilibrium \cite{deGroot}.
On the other side, phenomenological extensions to far-from-equilibrium
regimes \cite{Jou} seem to lack microscopic foundations.
For such systems, statistical temperatures have been proposed
in specific models, from FDR \cite{Puglisi}, maximum entropy principle
\cite{MaxEnt}, or generalized microcanonical approaches \cite{BDD}.
Yet, most of these studies do not discuss the relevance and generality of
the so-defined temperature.

In this Letter, we propose a general definition of ITPs conjugated to
additive conserved quantities in nonequilibrium systems, valid under
few assumptions. This definition, which does not require any detailed balance
relation to be fulfilled, is illustrated on a stochastic model in which
a dissipated energy is coupled to a globally conserved mass.
The usefulness of nonequilibrium ITPs to characterize
the coexistence of nonequilibrium phases is outlined, on the example
of a two-species mass transport model.
Finally, we discuss the problem of the contact between two nonequilibrium
systems with different microscopic dynamics, as well as the relation with
the large deviation functional approach \cite{Derrida02}.

{\it Framework and hypotheses.}--
Starting with a general formulation of the problem, we consider the
steady state of a macroscopic nonequilibrium system for which some additive
quantities $Q_k$, $k=1,\ldots,\ell$, are globally conserved by the dynamics.
On general grounds,
the stationary probability to be in a microstate $\alpha$, with probability
weight $f_{\alpha}$, is of the form
\be \label{dist-mucan}
P_{\alpha} = \frac{f_{\alpha}}{Z(Q_1,...,Q_{\ell})}
\prod_{k=1}^{\ell} \delta(Q_k^{\alpha}-Q_k)
\ee
where the product of delta functions accounts for the conservation laws,
and $Z(Q_1,...,Q_{\ell})$ is a normalization factor,
named hereafter a ``partition function''.
Splitting the system into two subsystems $\mathcal{S}_1$ and $\mathcal{S}_2$,
so that $\alpha=\{\alpha_1,\alpha_2\}$, we introduce the probability
$\Psi(Q_{11},\ldots,Q_{\ell 1}|Q_1,\ldots,Q_{\ell})$ that
conserved quantities have values $Q_{k1}$ in $\mathcal{S}_1$,
given their total values $Q_k$
\be
\Psi(\{Q_{k1}\}|\{Q_k\}) =
\sum_{\alpha_1,\alpha_2} P_{\alpha_1,\alpha_2}
\prod_{k=1}^{\ell} \delta(Q_{k1}^{\alpha_1}-Q_{k1}).
\ee
The key assumption in the following derivation is that this conditional
probability satisfies an asymptotic factorization property, namely
\bea \nonumber
\ln \Psi(\{Q_{k1}\}|\{Q_k\}) &=& \ln Z_1(\{Q_{k1}\}) + \ln Z_2(\{Q_k-Q_{k1}\})
\\
&-& \ln Z(\{Q_k\}) + \epsilon_N(\{Q_{k1}\},\{Q_k\})
\label{factorization}
\eea
where $Z_{\nu}(\{Q_{k\nu}\})$ refers to subsystem $\mathcal{S}_{\nu}$,
when isolated, and $\epsilon_N(\{Q_{k1}\},\{Q_k\})$ becomes negligible
with respect to $\ln Z_{\nu}$ in the large $N$ limit, $N$ being the number of
degrees of freedom.
Although not obvious in nonequilibrium systems where long-range
correlations might develop, this factorization property
actually holds in some rather large classes of models.
It is satisfied, for instance, in lattice models like
the ZRP \cite{Evans-Rev05}, the ARAP
\cite{ARAP}, or more general mass transport models
\cite{Zia04,Zia05}, where the probability weight
$f_{\alpha}$ factorizes as a product of one-site weights
$f_{\alpha}=\prod_{i=1}^N f_{i,\alpha_i}$, with $\alpha=\{\alpha_i\}$.
In this case, $\epsilon_N=0$ in Eq.~(\ref{factorization}).
Besides, this asymptotic factorization property also holds in
one-dimensional models with conservation laws, for which the stationary
probability is given by a matrix product ansatz with finite-dimensional
matrices: $f_{\alpha}=\mathrm{Tr} \prod_{i=1}^N M_{\alpha_i}$,
where $M_{\alpha_i}$ is a matrix associated to the local state $\alpha_i$
(for nonperiodic systems, one rather has 
$f_{\alpha}=\langle W| \prod_{i=1}^N M_{\alpha_i} |V\rangle$, with vectors
$\langle W|$ and $|V\rangle$ determined by the boundary conditions).
This approach has been used, e.g., in
the context of the ASEP model, and can lead to
finite-dimensional \cite{Mallick,Arndt} or infinite-dimensional matrices
\cite{Derrida93,Arndt}.
We shall simply sketch here the argument leading to Eq.~(\ref{factorization}),
and defer a more detailed presentation to a later publication \cite{long}.
For simplicity, we consider a generic model with local variables $q_i$
and one conserved quantity $Q=\sum_{i=1}^N q_i$.
The stationary distribution $P(\{q_i\})$ reads
\be
P(\{q_i\}) = \frac{1}{Z(Q)} \left( \mathrm{Tr} \prod_{i=1}^N M(q_i)
\right)\, \delta\left(\sum_{i=1}^N q_i -Q\right),
\ee
where $M(q)$ is a finite-dimensional matrix. Introducing
the matrix $R(Q)$ such that $Z(Q)=\mathrm{Tr}\, R(Q)$, one has
\be
\Psi(Q_1|Q) = \frac{1}{Z(Q)} \, \mathrm{Tr} [R_1(Q_1) R_2(Q-Q_1)]
\ee
Denoting by $\hat{R}(s)$ and $\hat{M}(s)$ the Laplace transforms
of $R(Q)$ and $M(q)$, one has $\hat{R}(s)=\hat{M}(s)^N$.
If $\hat{M}(s)$ is invertible, $\hat{M}(s)=\exp(B(s))$,
so that $\hat{R}(s)=\exp(NB(s))$.
$B(s)$ may be decomposed as $B(s)=D(s)+L(s)$ where
$D(s)$ is diagonalizable, $L(s)$ is nilpotent (i.e.~$L(s)^{p(s)}=0$),
and $D(s)$ commutes with $L(s)$.
It follows that $\hat{R}(s)=\exp(ND(s))\exp(NL(s))$,
where $\exp(NL(s))$ is a polynomial in $NL(s)$:
the dominant contribution to $\hat{R}(s)$ is proportional
to $N^{p(s)-1} \exp(N\ell_1(s))$, with $\ell_1(s)$ the largest
eigenvalue of $D(s)$.
If $p(s)$ is bounded when $s \to 0$, one can show performing an inverse
Laplace transform that the asymptotic factorization property
(\ref{factorization}) holds \cite{long}.

For infinite matrix representations, this argument
may break down as the eigenvalues of $D(s)$ might not be bounded.
From a physical viewpoint, it is reasonable to think that the factorization
property (\ref{factorization}) is related to the presence of a finite
correlation length in the system, as large subsystems then become
essentially independent. 
Both properties are expected to hold for finite matrices \cite{Arndt}.
Yet, for infinite matrices, whether these two properties are actually
related or not remains an open issue.

{\it Definition of nonequilibrium ITPs.}--
To define a nonequilibrium ITP, we compute, guided by equilibrium
procedures, the derivative of $\ln \Psi(\{Q_{k1}\}|\{Q_k\})$ with
respect to $Q_{k1}$ at the most probable value $Q_{k1}^*$.
Equating this derivative to zero yields
\be \label{equil-ITP}
\frac{\partial \ln Z_1}{\partial Q_{k1}} \Big\vert_{Q_{k1}^*} =
\frac{\partial \ln Z_2}{\partial Q_{k2}} \Big\vert_{Q_k-Q_{k1}^*}
\ee
Thus it is natural to define the ITP $\lambda_k$ of the system as
\be \label{def-ITP}
\lambda_k \equiv \frac{\partial \ln Z}{\partial Q_k}
\ee
since this parameter, when defined within subsystems that can exchange
the quantity $Q_k$, takes equal values in both subsystems
once the steady-state is reached, as seen from Eq.~(\ref{equil-ITP})
--see \cite{BDD} for a more detailed discussion on a specific example.
In the following, this property is loosely called the ``equilibration''
of subsystems, although we deal with nonequilibrium situations.

The above partition into two subsystems is also useful
when one subsystem is small, but still macroscopic.
The effect of the rest of the system, acting as a reservoir of $Q_k$,
may then be encoded into the parameters $\lambda_k$.
Integrating the distribution (\ref{dist-mucan}) over the degrees of
freedom of the reservoir leads for the subsystem to
\be \label{dist-can}
\tilde{P}_{\alpha} = \frac{f_{\alpha}}{\tilde{Z}(\lambda_1,...,
\lambda_{\ell})} \,\exp\left(-\sum_{k=1}^{\ell} \lambda_k Q_k^{\alpha}\right)
\ee
which may be called a ``grand-canonical'' distribution. Note also that
$\ln \tilde{Z}$ may be interpreted as a nonequilibrium thermodynamic
potential, and that, as in the equilibrium formalism, the cumulents of $Q_k$
are given by the successive derivatives of $\ln \tilde{Z}$ with respect
to $\lambda_k$ \cite{long}.

{\it Illustration on a dissipative model.}--
Let us illustrate on a simple nonequilibrium model the definition
(\ref{def-ITP}) of the ITP.
We consider a one-dimensional model with two different physical
quantities, say, an ``energy'' $E=\sum_{i=1}^N \ve_i$ and a ``mass''
$M=\sum_{i=1}^N m_i$. Energy is flowing through the boundaries, and is
partially dissipated in the bulk, whereas mass is conserved and cannot
flow through the boundaries.
The continuous time dynamics is defined as follows.
An amount of energy $\omega$ may be either injected at site $i=1$ with
rate $J(\omega)$, moved from site $i$ to $i+1$ with rate
$\varphi_1(\omega|\ve_i,m_i)$ (if $i=N$, the energy flows out of the system),
or dissipated on site $i$ with rate $\Delta(\omega|\ve_i,m_i)$.
Besides, an amount of mass $\mu$ may be transported from site $i$ to
$j=i\pm 1$ (except for $i=1$ or $N$, where $j=2$ or $N-1$ respectively)
with rate $\varphi_2(\mu|\ve_i,m_i)$, and with equal probability for both
target sites. To our knowledge, this model, which generalizes the cascade
model introduced in \cite{Bertin05}, was not considered previously in the
literature. In order to allow for a factorized distribution, the rate
functions are chosen as follows
\bea
\label{eq-J}
J(\omega) &=& v_1(\omega) e^{-b\omega}\\
\label{eq-phi1}
\varphi_1(\omega|\ve,m) &=& v_1(\omega)\, \frac{g(\ve-\omega,m)}{g(\ve,m)}\\
\label{eq-phi2}
\varphi_2(\mu|\ve,m) &=& v_2(\mu)\, \frac{g(\ve,m-\mu)}{g(\ve,m)}\\
\label{eq-Delta}
\Delta(\omega|\ve,m) &=& (e^{a\omega/N}-1)\, v_1(\omega)\,
\frac{g(\ve-\omega,m)}{g(\ve,m)}
\eea
with positive functions $v_1(\omega)$, $v_2(\mu)$, $g(\ve,m)$, and
two positive parameters $a$ and $b$.
Clearly, due to the presence of flux and dissipation, the above dynamics
cannot satisfy detailed balance. The steady-state distribution reads
\be
P(\{\ve_j,m_j\}) = \frac{\prod_{j=1}^N f_j(\ve_j,m_j)}{Z(M)} \;
\delta\left( \sum_{j=1}^N m_j-M \right)
\ee
with $f_j(\ve_j,m_j) = g(\ve_j,m_j)\, \exp[-(aj/N+b)\ve_j]$ on site $j$.
Let us choose for the function $g(\ve,m)$ the simple form
$g(\ve,m) = \exp(-\kappa m \ve^{\sigma})$ with $\sigma>1$ and $\kappa>0$.
The partition function can be computed in the large $N$ limit,
and one finds $Z(M) = A M^{(1-1/\sigma)N-1}$
with $A$ independent of $M$. This gives for the ITP
$\lambda = (\sigma-1)/(\sigma\rho)$
with $\rho=M/N$, showing that $\lambda$ depends, through $\sigma$,
on the coupling to the nonconserved energy.

{\it Relevance of ITPs for phase coexistence.}--
Let us now discuss the case
where two systems with the same microscopic dynamics, but different
macroscopic states, can exchange some globally conserved quantities.
A well-studied example of such a situation is the
condensation transition observed in ZRP
\cite{Spohn,Godreche,Hanney,Evans-Rev05}, or more general mass transport
models \cite{Zia05}. This transition occurs
when the overall density exceeds a critical value $\rho^{crit}$, so
that a finite fraction of the total mass condense onto a given site.
In standard approaches \cite{Evans-Rev05},
the critical density is obtained by looking for
the convergence radius of a formal grand-canonical partition function
\footnote{By ``formal'', we mean that it is defined by analogy with
equilibrium as a superposition of different canonical partitions functions,
without being clearly associated with a well-defined physical ensemble
--see, e.g., \cite{Arndt}.}.
Details about the condensed phase need to be studied in
the canonical ensemble where the total mass is fixed \cite{Zia05}.

In the following, we interpret the condensation as a phase coexistence
and show that enforcing the equality of the ITP defined in Eq.~(\ref{def-ITP})
--here, a chemical potential-- in the two phases
leads to a natural quantitative description of the condensation.
To this aim, we compute separately
the chemical potential of the condensate, taken as isolated,
and the equation of state of the fluid phase.
Let us illustrate this point on a concrete example.
We consider a variant of the model defined by Eqs.~(\ref{eq-J})
to (\ref{eq-Delta}), defined on a periodic lattice, where $\ve_i$ is
now a conserved quantity (that is,
$J(\omega)=0$ and $\Delta(\omega|\ve,m)=0$),
so that we shall replace the notations $\ve_i$ and $m_i$ with
$m_{1i}$ and $m_{2i}$ respectively (both quantities are now ``masses'').
The continuous time dynamics
proceeds by transferring a mass $\mu_1$ or $\mu_2$ from a
random site $i$ to $i+1$, with rates $\varphi_1(\mu_1|m_{1i},m_{2i})$
or $\varphi_2(\mu_2|m_{1i},m_{2i})$ given in Eqs.~(\ref{eq-phi1}) and
(\ref{eq-phi2}) respectively. In dimension $d>1$, the mass $\mu_1$ or $\mu_2$
may also be transferred in any direction transverse to the flux, with
equal probability.
The steady-state distribution in arbitrary dimension $d$ is given by
\be
P(\{m_{ki}\}) = \frac{\prod_{i=1}^N g(m_{1i},m_{2i})}{Z(M_1,M_2)}
\prod_{k=1}^2 \delta \left( \sum_{i=1}^N m_{ki} - M_k \right)
\ee
Guided by the two-species ZRP studied in \cite{Hanney}, we choose
$g(m_1,m_2) = \exp(-\kappa\, m_2\, m_1^{-\nu})$, with $\nu>0$.
Assuming that a condensate containing large amounts $M_{1c}$ and $M_{2c}$
forms, the partition function 
of the isolated condensate is $Z_c(M_{1c},M_{2c})=g(M_{1c},M_{2c})$, and
the chemical potentials $\lambda_k^c = \partial \ln Z_c/\partial M_{kc}$ read
\be \label{equal-lambda}
\lambda_1^c = \kappa \nu \frac{M_{2c}}{M_{1c}^{\nu+1}}, \qquad
\lambda_2^c = -\frac{\kappa}{M_{1c}^{\nu}}
\ee
Thus in the thermodynamic limit, $\lambda_2^c=0$.
Enforcing the equality of the chemical potentials in both phases
$\lambda_k^f = \lambda_k^c=\lambda_k$, the densities $\rho_k^f$ in the fluid
phase are computed in the grand-canonical ensemble, yielding
$\rho_1^f=(\nu+1)/\lambda_1$ and
$\rho_2^f=\Gamma(1+2\nu)/[\Gamma(1+\nu)\kappa \lambda_1^{\nu}]$.
Hence $\lambda_1 > 0$, and $M_{1c} \sim M_{2c}^{1/(\nu+1)}$.
The condensate $M_{1c}$ is subextensive, and $\rho_1^f=\rho_1$.
It follows that $\lambda_1$, and thus $\rho_2^f$, are fixed by the overall
density $\rho_1$.
A condensate forms if $\rho_2 > \rho_2^f$, meaning that $\rho_2^f$ is the
critical density $\rho_2^{crit}$. One has $M_{2c}=N(\rho_2-\rho_2^{crit})$,
and Eq.~(\ref{equal-lambda}) for $\lambda_1^c$ gives
\be
M_{1c} = \left[ \frac{\kappa\nu \rho_1}{\nu+1} N (\rho_2-\rho_2^{crit})
\right]^{\frac{1}{\nu+1}}
\ee
Accordingly, the framework of ITPs provides a simple description of the
condensation in terms of the (out-of-equilibrium) ``equilibration'' of two
coexisting phases.

{\it Contact of systems with different dynamics.}--
In this part, we consider the issue of the equalization of ITPs when
two systems with different microscopic dynamics are put into contact.
This is precisely the type of problem encountered when one wishes to define a
``thermometer'', which in the present more general nonequilibrium context,
we shall call ``ITP-meter''.
For such an ``ITP-meter'', one requires two essential properties.
First, its ITP must equalize with that of the system over which the measure
is performed, without perturbing this system.
Second, one needs to know the equation of state of the ``ITP-meter'' to relate
its ITP to a directly measurable quantity.
This is a highly non trivial problem, which depends on the
microscopic dynamics of each system, but also on the dynamics
at the contact.
Let us imagine that two different systems $\mathcal{S}_1$ and $\mathcal{S}_2$,
that separately conserve
the same physical quantity $Q=\sum_i q_i$, are put into contact.
The dynamics at the contact indeed imposes the distribution $\Phi(Q_1|Q)$
for the random partition of $Q$ into $Q_1$ and $Q_2=Q-Q_1$ over the two
systems. The probability distribution of the whole system then reads,
assuming for simplicity that the probability weights of each system strictly
factorize (the index $i_{\nu}$ refers to $\mathcal{S}_{\nu}$)
\be
P(\{q_i\}) = \frac{\Phi(Q_1|Q) \prod_{i_1,i_2}
f_1(q_{i_1}) f_2(q_{i_2})}{Z_1(Q_1) Z_2(Q-Q_1)}
\, \delta \left(\sum_i q_i-Q \right)
\ee
with $Q_1 \equiv \sum_{i_1\in \mathcal{S}_1} q_{i_1}$.
The ITPs of the two systems equalize only if
$\Phi(Q_1|Q)=Z_1(Q_1) Z_2(Q-Q_1)/Z(Q)$. This relation is indeed satisfied in
equilibrium due to the detailed balance relation, if the Hamiltonian is
additive \cite{long}.
Out of equilibrium, the above factorization and the equalization of the ITPs
does not hold in general, but are recovered inside some classes of
nonequilibrium systems.

To illustrate this point, let us consider two single conserved mass
transport models $S_1$ and $S_2$ with site-independent transport rates
$\varphi_{1,2}(\mu|m)=v(\mu)f_{1,2}(m-\mu)/f_{1,2}(m)$, with different
$f_1(m)$ and $f_2(m)$, but the same $v(\mu)$.
The dynamics at the contact is defined as follows: a mass leaving a boundary
site may be transferred to one of the neighboring sites or, with the same
probability, to the other system.
The global probability weights factorize (as $v(\mu)$ is
site-independent \cite{Zia04,long}), and the ITPs of the two systems equalize.
The key point is that the equalization of the ITPs occurs for two different
microscopic dynamics ($f_1(m)\neq f_2(m)$), but belonging to a given class
(same $v(\mu)$). Indeed, if $v_1(\mu)\neq v_2(\mu)$, such an equalization does
not hold. This result suggests the existence of classes of systems which
mutually ``equilibrate'', so that any member of a class, with known equation
of state, may be used as an ``ITP-meter'' for the other members, under
suitable size ratios.

{\it Discussion.}--
We have proposed a general definition of ITPs associated to
additive conserved quantities in nonequilibrium steady states,
under the asymptotic factorization assumption (\ref{factorization}).
This property holds in particular if the stationary distribution is
factorized or, for one-dimensional systems, if it can be computed
from a matrix product ansatz with finite-size matrices.
Its applicability to more general situations remains an important
open issue. Indeed, it would be essential to have a criterion allowing
to determine experimentally or numerically whether the
factorization property holds, without having to know the stationary
probability distribution explicitely.
Note that in equilibrium, the same difficulty
arises in principle, but it reduces there to the ``additivity'' of the
hamiltonian (short-range interactions).

Besides, we note that if $\Psi(Q_1|Q)$ may be written as
$\Psi(Q_1|Q) = \exp[-NG(\rho_1,\rho)]$, with $\rho_1=Q_1/N_1$
and $\rho=Q/N$, Eq.~(\ref{factorization}) corresponds to an additivity
property for $G(\rho_1,\rho)$, which somehow resembles the
additivity principle for the large deviation functional
$\mathcal{F}(\{\rho(x)\})$ of the density profile $\rho(x)$
in the open ASEP model \cite{Derrida02}.
As the physically relevant states minimize $G(\rho_1,\rho)$ with respect
to $\rho_1$ in the thermodynamic limit, $G(\rho_1,\rho)$
appears as a ``free energy'' function
\footnote{Introducing a partition into a large number of subsystems, one
may also construct in this way a large deviation functional for the
density profile $\rho(x)$.}. In analogy with the
ASEP where $\mathcal{F}(\{\rho(x)\})$ is a nonlocal functional
of $\rho(x)$, $G(\rho_1,\rho)$ is a nonlocal function as it depends on the
global density $\rho$, due to the conservation law. Still, the physical
origin of this nonlocality may be different in the two problems.
Also, the quantity $Q$ is conserved here,
whereas particles are exchanged with boundary reservoirs in the ASEP. 

From Eq.~(\ref{equil-ITP}), one sees that
the equality of ITPs between subsystems is related to
the minimization of a nonlocal ``free energy'' function.
Note however that this is already true in equilibrium systems.
The specificity of nonequilibrium dynamics shows up mostly when the systems
into contact have different microscopic dynamics.
In this case, the ITPs of the two systems may not equalize,
due to the essential role played by the dynamics at
the contact --a role hidden at equilibrium due to detailed balance.
Yet, there may exist classes of nonequilibrium systems that mutually
``equilibrate''.

{\it Acknowledgements.}--
This work has been partly supported by the Swiss National Science
Foundation.

\end{document}